\documentclass[a4paper,12pt]{article}
\usepackage{epsf,amsmath}
\usepackage{indentfirst}
\usepackage{amssymb}
\usepackage{hhline}
\usepackage{textcomp}
\usepackage{array,tabularx}
\def\@biblabel#1{#1.}
\makeatother

\begin{document}

\title{\Large\bf A Study of the Evolution of the Close Binaries Cyg X-3,
IC 10 X-1, NGC 300 X-1, SS 433, and M33 X-7 Using the ``Scenario Machine''}

\author{A. I. Bogomazov\\
\small\it M. V. Lomonosov Moscow State University, \\
\small\it P. K. Sternberg Astronomical Institute, \\ Moscow, Russia }

\date{\begin{minipage}{15.5cm} \small
Evolutionary tracks for the X-ray binaries Cyg X-3, IC 10 X-1, NGC 300 X-1, SS 433,
and M33 X-7 are computed using the Scenario Machine code. The compact objects in IC 10 X-1,
NGC 300 X-1, and M33 X-7 are the most massive stellar mass black hole candidates. Cyg X-3,
IC 10 X-1, and NGC 300 X-1 are the only currently known Wolf-Rayet stars with degenerate companions.
SS 433 is the only known superaccretor in the Milky Way. Therefore, the stars studied provide excellent
laboratories for testing scenarios for the evolution of binaries under extreme conditions. The classical
evolutionary scenario is consistent with modern observational data. During the evolution of these binaries,
hypernova explosions accompanied by the collapse of stellar cores with large angular momenta can occur,
leading to long gamma-ray bursts. At the end of their evolution, Cyg X-3, IC 10 X-1, NGC 300 X-1,
and SS 433 may form binary relativistic objects, which will subsequently merge due to the radiation
of gravitational waves. The gravitational waves emitted during mergers of relativistic stars should be
detectable by existing and future gravitational-wave antennas. In the course of its future evolution,
M33 X-7 will pass through a Thorne-Zytkow stage. The formation of a Thorne-Zytkow object can also
be accompanied by gravitational-wave radiation.
\end{minipage} } \maketitle \rm

\section{Introduction}

The possibility of forming Wolf-Rayet stars in
massive close binaries was considered in the classical
paper \cite{tutukov1973a}. Arguments suggesting the loss of mass
and angular momentum in the course of the mass
transfer preceding the formation of the Wolf-Rayet
star were also provided in \cite{tutukov1973a}. The evolution of such
systems can include a stage with a close binary composed
of a Wolf-Rayet star and a compact degenerate
stellar remnant \cite{tutukov1973b}. Currently, three X-ray binaries
harboring Wolf-Rayet stars and stellar mass black hole
candidates are known in the Milky Way and
other nearby galaxies: Cyg X-3, IC 10 X-1, and
NGC 300 X-1. Certain aspects of the evolution of
these three objects were studied in \cite{tutukov2013}, together with
the evolution of the close binary SS 433, which is a
possible precursor to a Wolf-Rayet + compact remnant
system, assuming that SS 433 will not undergo
a common-envelope stage in its subsequent evolution
(as it is currently in such a stage).

The aim of the present paper is to study the evolutionary
paths of the massive close binaries Cyg X-3,
IC 10 X-1, NGC 300 X-1, SS 433, and M33 X-7 using the ``Scenario Machine''\footnote{A code for population-synthesis studies of close binaries
\cite{lipunov1996,lipunov2009}.}. The choice of these
systems is not coincidental. There are only three
known black-hole candidates in binaries with Wolf-Rayet stars (Cyg X-3, IC 10 X-1, NGC 300 X-1),
and their evolutionary path should go through several
stages. SS 433 may be a precursor to a black hole +
Wolf-Rayet binary, and is currently the only known
superaccretor in the Milky Way. M33 X-7 is a close
binary harboring a black-hole candidate (one of the
most massive) and a slightly evolved non degenerate
star. These systems can be considered to be ``cornerstones''
of the theory of the evolution of close
binaries. Studies of their evolution enable verification
of evolutionary scenarios for extreme conditions.

The evolution of close binaries such as those studied
here can include the collapse of very massive
stellar cores. Since they are components of binary
systems, they can possess large angular momenta.
Such events are usually associated with long
gamma-ray burThe evolution of close binaries such as those studied
here can include the collapse of very massive
stellar cores. Since they are components of binary
systems, they can possess large angular momenta.
Such events are usually associated with long
gamma-ray bursts\footnote{See, for instance,
\cite{bogomazov2007,lipunova2009}, where the rate of Wolf-Rayet core
collapses in extremely close binary systems is studied using
the Scenario Machine.}. Depending on the parameters of the evolutionary scenario, the evolution may end in
the formation of a binary with two relativistic objects
or of two single objects, resulting from the disruption
of the system. If a close relativistic binary forms,
its components will merge due to gravitational-wave
radiation. Such a merger would produce an outburst
of gravitational waves that should be detected by LIGO\footnote{See, for instance,
\cite{lipunov1997}, where the rate of relativistic binary
mergers is studied using the Scenario Machine.}. If one or both components of such a system
is a neutron star, the merger will also produce a short
gamma-ray burst.

Thus, studies of the evolution of Cyg X-3,
IC 10 X-1, NGC 300 X-1, SS 433, and M33 X-7
are topical, since they enable the identification of the
parameters of evolutionary scenarios and possible
sources of gravitational waves and gamma-ray bursts
(both long and short).

We will now briefly describe the systems studed.
Cyg X-3 and SS 433 are Galactic objects, while
IC 10 X-1, NGC 300 X-1, and M33 X-7 are located
in the galaxies IC 10, NGC 300, and M33, respectively.

{\bf Cyg X-3} is a Wolf-Rayet star \cite{kerkwijk1996a} paired with a
compact object, which is believed to be a black hole
(see, e.g., the review \cite{cherepashchuk2003a}). The orbital period of the
system is 4.8 hr. In early studies, the compact object
was usually taken to be a black hole with a mass from $7M_{\odot}$ to
However, later studies yielded a
different result: the mass of the relativistic star does
not exceed \cite{hanson2000a}, admitting the presence of
either a neutron star or a black hole in the system.
According to the modern data on the Doppler shifts
of X-ray lines, the mass of the compact object in
Cyg X-3 does not exceed $3.6
M_{\odot}$ \cite{stark2003a}. This still
allows the possibility that this is a system with a black
hole paired with a Wolf-Rayet star, but does not
provide a final answer to its nature. An additional argument
in favour of the hypothesis that there is a black
hole in the system is that accreting neutron stars
paired with Wolf-Rayet stars probably cannot exist,
since their rotation would be strongly accelerated
during the second episode of mass transfer, so that
they would become either ejectors or propellers \cite{lipunov1982a}. The X-ray luminosity of the Cyg X-3 system at 1-60 keV is $\sim 10^{38}$ erg/s and tahe bolometric luminosity
of theWolf–Rayet star is $\approx 3\cdot 10^{39}$ erg/s.

{\bf IC 10 X-1} consists of a Wolf-Rayet star and one of
the most massive stellar black hole candidates \cite{prestwich2007}. The mass of the Wolf–Rayet star is $32.7\pm 2.6 M_{\odot}$, and the mass of the black hole estimated from the
observed radial-velocity curve of the optical component
is $23.1 \pm 2.1
M_{\odot}$ \cite{silverman2008a}. The orbital period of the
system is 34.4 hr. IC 10 X-1 is a bright, variable
X-ray source in the metal-poor galaxy IC 10, which
is underoing a burst of star formation. The X-ray
luminosity of the system is $\sim 10^{38}$ erg/s \cite{brandt1997,bauer2004}. The
most probable optical counterpart of the X-ray source
is the bright Wolf-Rayet star
[MAC92] 17A \cite{crowther2003}. The
nature of a synchrotron superbubble in IC 10 was
studied in
\cite{lozinskaya2007,lozinskaya2008}, where it was shown that the most
plausible mechanism for the formation of this bubble
is a hypernova explosion. IC 10 X-1 is located inside
this superbubble, and it is possible that formation of
the two objects is related.

{\bf NGC 300 X-1} became the third known Wolf-Rayet binary with a degenerate companion \cite{caprano2007}. The
orbital period of the system is $32.8\pm 0.4$ hr. According
to XMM Newton data, the mean observed
luminosity of the system at 0.2-10 keV is approximately $2\cdot 10^{38}$ erg/s, reaching
$\sim 10^{39}$ erg/s (taking into account absorption along the line of sight).
According to \cite{crowther2010} the radial-velocity semi-amplitude
is $267 \pm 8$ km/s, yielding a mass function of $2.6 \pm 0.3
M_{\odot}$, the orbital period is $32.3 \pm 0.2$ hr. The
spectroscopic mass of the optical component (WN
Wolf-Rayet star) is $26^{+7}_{-5} M_{\odot}$, implying amass for the
black hole of $20 \pm 4 M_{\odot}$ for the most likely inclination
of the orbit, 60-75$^{\circ}$. If only half of the optical
continuum emission is provided by the Wolf-Rayet
star, its mass is $15^{+4}_{-2.5} M_{\odot}$, and the mass of the black
hole is $14.5^{+3}_{-2.5}M_{\odot}$.

{\bf SS 433} is the only known supercritical accretor
in the Milky Way. Although this system has been
known for a long time, and has been the subject
of hundreds of papers, there is still ample scope for
further studies. SS 433 is a close, eclipsing binary
with an orbital period of about 13 days (see, e.g., the
review \cite{fabrika2004}), in which the donor fills its Roche lobe
and material overflows onto a relativistic component
(black-hole candidate) on the thermal time scale, at
an accretion rate of $\sim 10^{-4} M_{\odot}$/yr. Estimates of the
mass of the compact object range between 2 and
15 $M_{\odot}$ \cite{fabrika1990}--\cite{kubota2010}, and do not enable unambiguous
conclusions about the nature of the degenerate component.

{\bf M33 X-7} was discovered in the early 1980s
\cite{long1981}. Periodic variability of the X-ray source was discovered
in \cite{peres1989a,peres1989b} and it was hypothesized that the
source is one component of a binary. Later, an O6 III
optical counterpart of M33 X-7 was found, which has
minimum mass of $20 M_{\odot}$ and an orbital period of $3.45^{d}$ \cite{pietsch2004,pietsch2006}. Analysis of the observed radial-velocity
curve of the X-ray binary using a Roche model and
of the dependence of the component masses on the
Roche-lobe filling factor was performed in
\cite{abubekerov2009}. For
themost probablemass of the optical star, $70 M_{\odot}$, the
mass of the compact object was found to be $15.55 \pm 3.20
M_{\odot}$, placing it among the most massive stellar
black-hole candidates.

The evolution of IC 10 X-1 and M33 X-7 has
already been studied using the Scenario Machine in \cite{abubekerov2009}, where the most characteristic evolutionary
tracks were constructed. In our present study,
we carry out additional studies of IC 10 X-1 and
M33 X-7.

\section{Population synthesis}

Our studies of the origin and subsequent evolution
of the close binary systems Cyg X-3, IC 10 X-1,
NGC 300 X-1, SS 433, and M33 X-7 were carried
out using the ``Scenario Machine''. This program is
designed for studies of the evolution of close binaries
through population-synthesis techniques, which can
be applied to compute individual evolutionary tracks
of close binaries and investigate the properties of
groups of close binaries of various types. A detailed
description of the ``Scenario Machine'' is provided in
\cite{lipunov1996,lipunov2009}, and the method used for the close-binary population
synthesis is described, for instance, in the review
paper \cite{popov2007}. In our present study, we took the problem
to involve three free parameters: the rate of mass loss
by the non-degenerate star, the fraction of the mass
of the pre-supernova star that disappears beyond the
event horizon during the formation of the black hole,
and the efficiency of the common envelope stage. The
values of other parameters were not varied, and were
taken to be equal to their standard values during the
computations. For each set of initial parameters, a
population synthesis was performed for $10^6$ binary systems.

The stellar mass loss rate $\dot M$is very important for
two reasons: first, it significantly affects the semi-major
axis of the binary and, second, it directly affects
the mass of the star itself. We considered evolutionary
scenario A from \cite{lipunov2009} only, in which mass loss by
main sequence stars is described using the classical
formula

\begin{equation}
\dot M =\frac{\alpha L}{cV_{\infty}},
\end{equation}

\noindent where $L$ is the stellar luminosity, $V_{\infty}$ the stellarwind
velocity at infinity, $c$ the speed of light, and $\alpha$ a free parameter. In scenario A, the decrease in the
stellar mass $\Delta M$ does not exceed
$0.1(M-M_{core})$ during any one evolutionary stage, where $M$ is mass
of the star at the beginning of the stage and
$M_{core}$ is the mass of the stellar core. We parametrized
the mass-loss by Wolf-Rayet stars as
$\Delta M_{WR}= \alpha M_{WR}$, where $M_{WR}$ is the maximum stellar mass in
the Wolf-Rayet stage.

The mass of a black hole $M_{BH}$ formed by an
exploding pre-supernova of mass $M_{preSN}$, was calculated as

\begin{equation} M_{BH}=k_{bh}M_{preSN}, \label{k_BH} \end{equation}

\noindent where the coefficient $k_{bh}$ is the fraction of the presupernova
mass that disappears beyong the event
horizon during the collapse, which was varied between 0.1 and 1.0.

During a common envelope stage, stars very efficiently transfer their angular momentum to the surrounding matter, and approach one another along a spiral trajectory. The efficiency of mass loss in the common envelope stage is described by the parameter $\alpha_{CE}=\Delta E_b/\Delta E_{orb}$, where $\Delta
E_b=E_{grav}-E_{thermal}$ is the binding energy of the ejected envelope
and $\Delta E_{orb}$ is the reduction in the orbital separation
during the approach:

 \begin{equation} \alpha_{CE} \left( \frac{GM_a M_c}{2a_f} - \frac{GM_a M_d}{2a_i} \right)=\frac{GM_d(M_d - M_c)}{R_d}, \end{equation}

\noindent where $M_c$ is the core mass of the mass losing star,
which has initial mass $M_d$ and radius $R_d$ (this is a function of the initial semimajor axis $a_i$ and initial
component-mass ratio $M_a/M_d$, where $M_a$ is the
mass of the accretor).

In our computations, Cyg X-3, IC 10 X-1, and
NGC 300 X-1 were taken to be black holes with
Wolf-Rayet companions. Since the mass of the
degenerate star is not known well, we considered
two options: a low mass and a massive degenerate
star. In the low mass version, the mass of the black
hole was $\le 10 M_{\odot}$,
the mass of the Wolf-Rayet was $\le 10 M_{\odot}$, the the orbital period of the system was $\le 0.2$ day. In the high mass version, the mass of the
black hole was $\ge 10 M_{\odot}$, the mass of the Wolf-Rayet
star was  $\ge 10
M_{\odot}$, the orbital period was $\le 0.2$ day,
and the X-ray luminosity was $\ge 10^{38}$ erg/s. We
adopted the following parameters for IC 10 X-1, in
accordance with available observational data: mass
of the black hole $23\div 34 M_{\odot}$, mass of the Wolf-Rayet
star $17\div 35 M_{\odot}$, and the orbital period $\le
1.5$ day. For NGC 300 X-1, we adopted the mass of the black hole
to be $20\div 25
M_{\odot}$, orbital period to be $\le 1.5$ day, and
X ray luminosity to be $\ge
10^{38}$ erg/s; the mass of the
Wolf-Rayet star was not constrained.

SS 433 was taken to be a black hole undergoing
supercritical accretion paired with a non-degenerate
star that overflows its Roche lobe. The mass of the
optical star was taken to be $5\div 15 M_{\odot}$.

M33 X-7 was taken to be a $14\div 17 M_{\odot}$ black hole
paired with a $65\div
75 M_{\odot}$ star that is close to the end
of its main-sequence evolution. The orbital period of
the system was limited to 5 days.

\section{Results of the computations}

\label{results}

Results of the computations are presented in
Figs. \ref{cygss}-\ref{ris-ic10x1-ns} and Tables \ref{tr1}-\ref{tr3}.
The main aim of our studies using the ``Scenario Machine'' is to determine
the most probable range of parameters of evolutionary
scenarios describing these systems. For this, we
constructed a plot (Fig. \ref{cygss})
of the number of Cyg X-3 
type and NGC 300 X-1 type systems (for the case
of a massive black hole) in an
$10^{11}M_{\odot}$ spiral
galaxy with a Salpeter star formation rate as
a function of the fraction of the mass of the presupernova
that disappears beyond the event horizon
during the formation of the black hole $k_{bh}$. The
curves in Fig. \ref{cygss} were computed for various efficiencies
of the common-envelope stage $\alpha_{CE}$and mass loss
rates $\alpha$. Since each of the systems considered is the
only object of its type in its parent galaxy, the main
criterion for selecting a suitable set of evolutionary
parameters was that the number of such systems be
roughly equal to 1 in the model. If possible, the set
of parameters satisfying all systems simultaneously
was chosen. Figure \ref{cygss} shows that this condition is satisfied  for Cyg X-3 and NGC 300 X-1 for
$\alpha_{CE}=0.5$, $a=0.3$, and $k_{bh}=0.5$. This set of parameters
was used to compute the evolutionary tracks of the
systems considered\footnote{Similar plots were constructed for M33 X-7 type and IC 10 X-1 type in \cite{abubekerov2009}, Figs. 8 and 10.}.

The following notation is used for the evolutionary
stages in Figs. \ref{ris-cyg-x-3-massive}-\ref{ris-ic10x1-ns} and Tables \ref{tr1}-\ref{tr3} ïðèíÿòû ñëåäóþùèå îáîçíà÷åíèÿ
ýâîëþöèîííûõ ñòàäèé (ñì. ðàáîòó \cite{lipunov2009}): I -- main sequence star, II -- supergiant, III, IIIe, IIIs --
Roche lobe filling star, WR -- Wolf-Rayet star, BB --
Roche lobe filling Wolf-Rayet star, BH -- black hole, SBH --
superaccreting black hole, Psr --
radio pulsar, CE -- common envelope, and SN -- supernova explosion.
The parameters presented in the figures and tables are: $M_1$ and $M_2$ -- masses of primary and secondary
components (in solar units), $a$ -- semimajor axis of
the orbit in $R_{\odot}$, $T$ -- time elapsed since the formation
of the system (in Myr), and $e$ -- orbital eccentricity. All values of $M_1$, $M_2$, $a$, $T$, and $e$ are listed for the beginning of the corresponding stage; the time given
for a supernovae is the time just before the explosion.

Figure \ref{ris-cyg-x-3-massive} presents a characteristic evolutionary
track for a binary resulting in the formation of a Cyg X-3 type system (massive black hole). At
the beginning of the evolution, the mass of the primary (more massive) star is $M_1 = 60\div 120 M_{\odot}$, the component-mass ratio is $q=\frac{M_2}{M_1}=0.3\div 0.8$, and the initial semi-major axis of the orbit is $a=70\div 140 R_{\odot}$. Note that the mass of a star can increase
during mass transfer, so that the initial mass of a
star may be lower than its mass in a subsequent
evolutionary stage. After the depletion of hydrogen in
the core, the primary star fills its Roche lobe. Matter
begins to overflow onto the companion, as a rule,
faster than the nuclear time scale (Stage III); the
final phase of the overflow occurs on the nuclear time
scale (Stage IIIe). The remnant of the primary after
the loss of its envelope becomes a Wolf-Rayet star,
which explodes and forms a black hole. Further, the
secondary completes its main sequence evolution,
becomes a supergiant, and, subsequently also fills its
Roche lobe. In this configuration, the accretion rate
becomes supercritical. A common envelope rapidly
forms, the binary components become very close, and
the envelope of the non-relativistic component is lost.
A pair consisting of a black hole and a Wolf-Rayet
star forms. The binary becomes so close that the
Wolf-Rayet star may overflow its Roche lobe. During
the collapse of a Wolf-Rayet star in such a binary, a
long gamma-ray burst may occur. The final stage of
the evolution of the system in this case\footnote{We assumed that the black holes did not receive a significant
kick during their formation.} is the merger
of two black holes (the remnants of the components)
due to gravitational-wave radiation, resulting in the
formation of a single massive black hole. The merger
of two black holes should produce an outburst of
gravitational-wave radiation.

A characteristic evolutionary track leading to the
formation of a Cyg X-3-type system with a lowmass
black hole is presented in Fig. \ref{ris-cyg-x-3-low}. At the beginning
of the evolution, the mass of the primary
star is $M_1 = 25\div 37
M_{\odot}$, the component mass ratio
is $q=\frac{M_2}{M_1}=0.2\div 0.9$,
and the initial semimajor axis
of the orbit is between $a=30$ and $130 R_{\odot}$. After
the depletion of hydrogen in the core, the primary
becomes a supergiant and subsequently fills its Roche
lobe. The stellar material overflows from the lobe on a
time scale that is shorter than the nuclear time scale
(Stage III). After the loss of the envelope, the primary
becomes a Wolf-Rayet star, which leaves a black hole
remnant after its explosion as a supernova. Next,
the secondary finishes its main sequence evolution,
becomes a supergiant, and later overflows its Roche
lobe. The accretion rate becomes supercritical. A
common envelope rapidly forms, in which the components
become very close. The envelope of the non-relativistic
component is lost. A binary ``black hole +
Wolf-Rayet'' system forms. A long gamma ray burst
can also occur during the collapse of the Wolf-Rayet
star in such a binary. Since the component masses
are lower in this scenario than in the scenario described
above, as a rule, the remnant of the secondary
is a neutron star rather than of a black hole.

Depending on an additional parameter --- the kick
obtained by the neutron star when it is born, such
a system can produce either a single black hole and
a single neutron star (which will manifest itself as
a radio pulsar at the beginning of its lifetime), or a
binary system containing a black hole and neutron
star. In the latter case, it is possible that the pair is so
close that the black hole and neutron star will merge
due to the radiation of gravitational wave, producing
a short gamma-ray burst and an outburst of gravitational
waves. However, investigation of the influence
of the possible kick accompanying the birth of the
neutron star is beyond the scope of this paper.

Figure \ref{ris-ngc-300-x-1} presents a characteristic evolutionary
track of a binary leading to the formation of an NGC 300 X-1 type system. At the beginning of
the evolution, the mass of the primary is $M_1 = 90\div 120
M_{\odot}$, the component mass ratio is $q=\frac{M_2}{M_1}=0.35\div 0.75$,
and the initial semi-major axis of the orbit
is $a=70\div 160 R_{\odot}$. After hydrogen depletion in
the core, the primary star fills its Roche lobe. Mass
transfer onto the companion begins, which, as a rule,
proceeds on a time scale shorter than the nuclear
time scale (Stage III). The loss of the primary's
envelope results in the formation of a Wolf-Rayet
star, which forms a black hole when it explodes as a
supernova. Further, after finishing its main sequence
evolution, the secondary overflows its Roche lobe.
A common envelope rapidly forms, in which the
components become very close. The envelope of the
non-relativistic component is lost. A black hole and
Wolf-Rayet star in a pair is formed. The collapse of
the Wolf-Rayet star may be accompanied by a long
gamma ray burst. The result of the evolution of this
binary is the merger of two black holes remnants due
to gravitational wave radiation, with the formation
of a single massive black hole. The merger of two
black holes produces an outburst of gravitational wave radiation.

Figure \ref{ris-ic10x1-ns}presents an example of one track from a
family of possible evolutionary tracks of binaries that
result in the formation of IC 10 X-1-type systems \footnote{The track shown Fig. \ref{ris-ic10x1-ns}
supplements the track presented
in Fig. 11 in \cite{abubekerov2009}, where calculations were carried out with
a slightly different set of parameters, in particular, a lower
mass-loss rate.}. At the beginning of the evolution, the mass of the primary is $M_1 = 75\div 120 M_{\odot}$, the component mass ratio is
$q=\frac{M_2}{M_1}=0.15\div 0.6$, and the initial semimajor
axis of the orbit is $a=70\div 280 R_{\odot}$.
After depletion of hydrogen in the core, the primary overflows its
Roche lobe, and forms a common envelope around the
two non-degenerate stars. In this stage, the components
become very close, and the envelope of the
primary is partially lost and partially accreted by the
secondary. After the loss of the primary's envelope, a
Wolf–Rayet star forms. The binary becomes so close
that the secondary overflows its Roche lobe before the
depletion of hydrogen in its core. After the explosion
of the Wolf-Rayet primary and the collapse of its core
into a black hole, a binary system containing a black
hole and a star filling its Roche lobe forms. Accretion
proceeds on the nuclear time scale (Stage IIIe). After
the loss of its envelope, the secondary becomes a
Wolf-Rayet star, which subsequently explodes and
forms a neutron star. The system is already so close
at the time of the first supernova explosion that the formation of the black hole may be accompanied by a
gamma-ray burst\footnote{A gamma ray burst can probably occur during the collapse
of the core of the Wolf-Rayet star if the orbital period of the
binary is $\lesssim 0.5$ day \cite{lipunova2009}.}.

After the loss of the secodary's envelope, it likewise
becomes a Wolf-Rayet star. The explosion of the
secondary produces a neutron star, which is manifest
as a radio pulsar at the beginning its evolution. If
the kick obtained by the newborn neutron star is not
large, the resulting binary harbors a black hole and a
radio pulsar, which later fades. If the newborn neutron
star does not receive a significant kick, the system is
wide enough to avoid a merger of the components
over the Hubble time (this option is presented in
Fig. \ref{ris-ic10x1-ns}). A kick could also to disruption of the system
or a reduction in the semi-major axis of the orbit, with
the latter resulting in the merger of the components
due to gravitational-wave radiation over a reasonable
time. In the latter case, a short gamma ray burst
and an outburst of gravitational wave radiation could
occur.

Three possible and approximately equally probable
evolutionary tracks for the SS 433 system are
presented in Tables \ref{tr1}-\ref{tr3}. For a very wide range of
parameters in the evolutionary scenarios, the number
of SS 433-type systems in the Galaxy significantly
exceeds unity, and is equal to 10-20. Thus, other systems similar to SS 433 should exist in the Milky
Way\footnote{Uncertainty in the masses of the SS 433 components means
that we cannot definitively rule out a neutron star as the compact
component of the system. The observational manifestations
of a magnetized neutron star in a system with supercritical
accretion are similar to the properties of
SS 433 \cite{lipunov1982}, and the number of the binaries classified as ``super accreting
neutron star + Roche lobe filling non-degenerate star'' in
the Galaxy should also reach 20 \cite{lipunov1987}.}. The initial parameters of the precursors of
SS 433 type systems is poorly constrained. The
initial masses of the stars must be such that the
primary could form a black hole after its explosion
as a supernova (i.e., at some stage of its evolution
before the supernova, its mass must exceed 25 $M_{\odot}$),
and the initial semimajor axis of the orbit must enable
the overflow of the secondary's Roche lobe. The initial
component mass ratio can vary over a very wide
range: 0.1-0.9.

The evolution of the system most similar to
SS 433 is presented in the Table \ref{tr1}. The primary of
this very close binary fills its Roche lobe before the
depletion of hydrogen in its core. The time scale for
mass transfer onto the secondary is initially shorter
than the nuclear time scale (Stage III), but becomes
comparable to the nuclear time scale at the end of
the stage (Stage IIIe). After the loss of its envelope,
the primary becomes a Wolf-Rayet star. Further, the
secondary fills its Roche lobe before the completion
of hydrogen burning in its core. The time scale for
the mass transfer onto the primary is initially shorter
than the nuclear time scale, but then becomes similar
to the nuclear time scale. After the explosion of the
primary as a supernova, a black hole undergoing
supercritical accretion is formed. We identify this
stage with an SS 433 type star. The age of the
nebula surrounding SS 433, which was apparently
formed in a supernova explosion, W50, does not
exceed $10^5$ yr (see, for instance,
\cite{goodall2011} and references
therein). This makes the track presented in the Table \ref{tr1}, most plausible as a description of the evolutionary
path followed by SS 433. After the loss of its envelope
in the process of supercritical accretion, the optical
star becomes a Wolf-Rayet star. After the explosion
of the secondary, the system becomes a binary black
hole. The tracks presented in Tables \ref{tr2} and
\ref{tr3} exhibit
twomain differences compared to the track in the Table \ref{tr1}. First, after the first supernova explosion in the system, the secondary fills its Roche lobe several hundred
thousand years after the explosion as it leaves the
main sequence, which does not agree with the age
of W50. Second, a common envelope stage begins
after the SS 433 stage, supplementing the possible
paths for the subsequent evolution of the SS 433 
type systems considered in \cite{tutukov2013}.

Our study of the evolution of the M33 X-7 system
assuming various evolutionary parameters did not
reveal additional evolutionary scenarios compared to
those illustrated in Fig. 9 in
\cite{abubekerov2009}. This is true because
the parameters of M33 X-7 are known much more
accurately than those of the other systems considered
here. Therefore, the ranges of possible component
masses and orbital semimajor axes used in our calculations
were the narrowest for this system. Thus,
M33 X-7 stillmost tightly constrains the evolutionary
scenarios for the systems considered systems. In the
evolutionary models included in the Scenario Machine,
the evolutionary track for M33 X-7 has two
significant differences from the other tracks: first, it
is not possible to obtain an M33 X-7-type system
with appreciably higher mass loss rates by the non-degenerate
stars ($\alpha=0.3$,
$k_{WR}=0.3$), and, second,
the most probable fraction of pre-supernova mass
that disappears beyond the event horizon, $k_{bh}$, is less
than 0.5, and approximately equal to 0.3.

\section{Conclusions}

We have investigated possible evolutionary tracks
for the X-ray binaries Cyg X-3, IC 10 X-1,
NGC 300 X-1, SS 433, and M33 X-7 using the
``Scenario Machine''. Each of these systems can be
considered a potential source of gravitational-wave
radiation, and can potentially produce one to three
gamma ray bursts in the course of its evolution\footnote{Up to two long bursts and a short burst in the case of
IC 10 X-1. According to
\cite{lipunova2009} a gamma ray burst can occur
during the first supernova explosion in a system with the
parameters given in Fig. \ref{ris-ic10x1-ns},
whereas this is not possible
according to
\cite{heuvel2007} since the Wolf-Rayet star must be in a
binary system with a degenerate companion. In this case,
a gamma ray burst will occur only after the second supernova
explosion. However, the existence of a bubble around
IC 10 X-1 supports the possibility of a gamma-ray burst
already after the first supernova explosion \cite{lozinskaya2007,lozinskaya2008}.}. During the evolution of SS 433, after the stage corresponding to the current state of this system,
a binary harboring a Wolf-Rayet star and a black hole should form, with properties close to those of
Cyg X-3, NGC 300 X-1, and IC 10 X-1. In Figs. \ref{ris-cyg-x-3-massive}-\ref{ris-ic10x1-ns}, the evolutionary stages preceeding the formation of
these systems are also similar to SS 433, but have
higher masses. Unlike the results of \cite{tutukov2013}, SS 433 and the precursors of Cyg X-3, NGC 300 X-1, and
IC 10 X-1 can have a stage with common envelope after the stage corresponding to the current state of
SS 433. Some of the evolutionary tracks presented here include a ``radio pulsar + black hole'' stage.
The discovery of such a system would provide firm confirmation of the existence of stellar mass black holes \cite{lipunov2005}.

\begin{figure*}[h!]
\hspace{0cm} \epsfxsize=0.99\textwidth\centering
\epsfbox{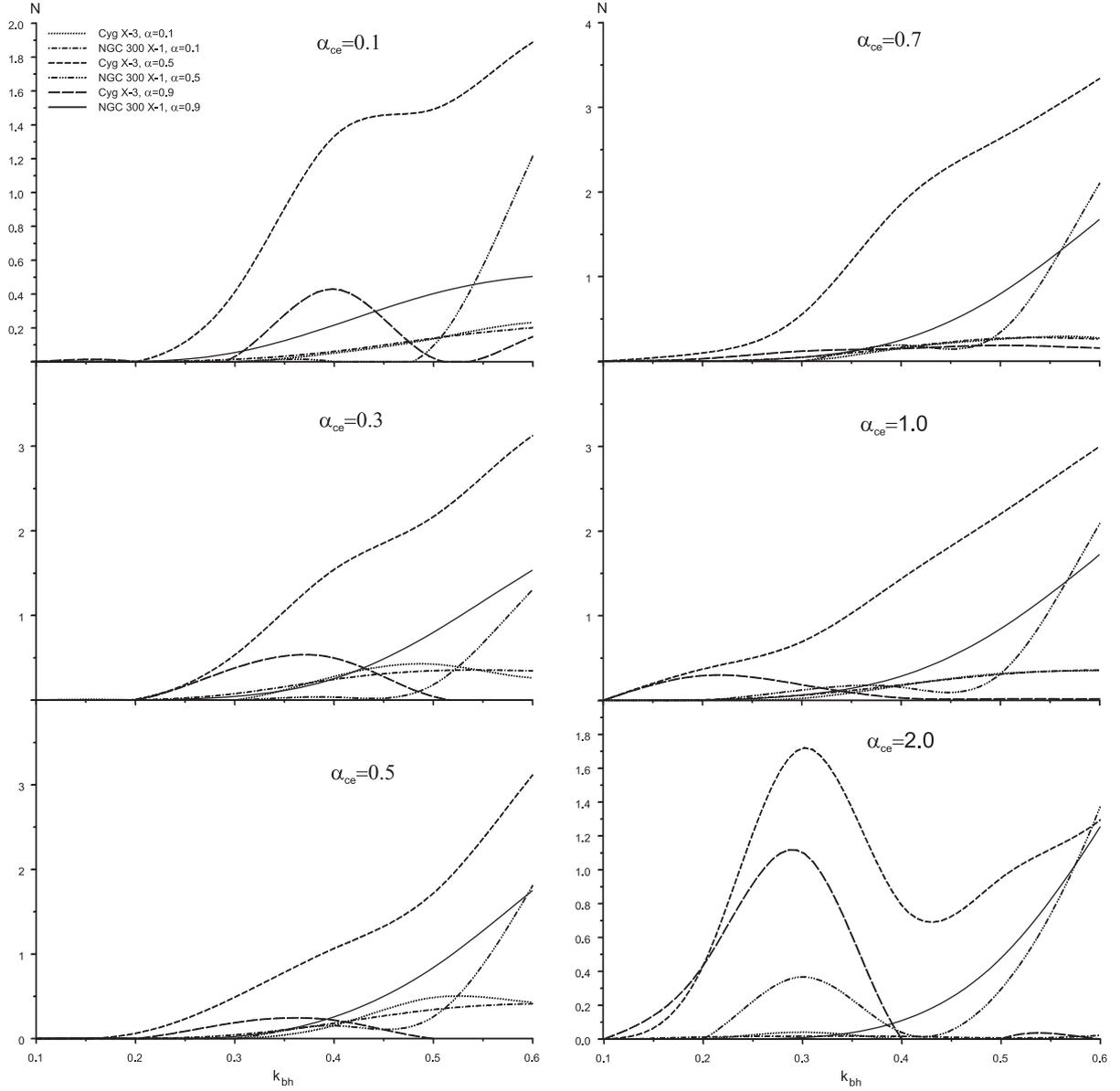} \vspace{0cm}\caption{Number of Cyg X-3 and NGC 300 X-1-type systems in a $10^{11}$
M$_{\odot}$ spiral galaxy with a Salpeter star formation rate as
a function of the fraction of the pre-supernova mass disappearing beyond the event horizon during the formation of the black hole, $k_{bh}$. The curves were computed for different efficiencies of the common envelope stage $\alpha_{CE}$ and mass loss rates $\alpha$.} \label{cygss}
\end{figure*}

\begin{figure*}[h!]
\hspace{0cm} \epsfxsize=0.95\textwidth\centering
\epsfbox{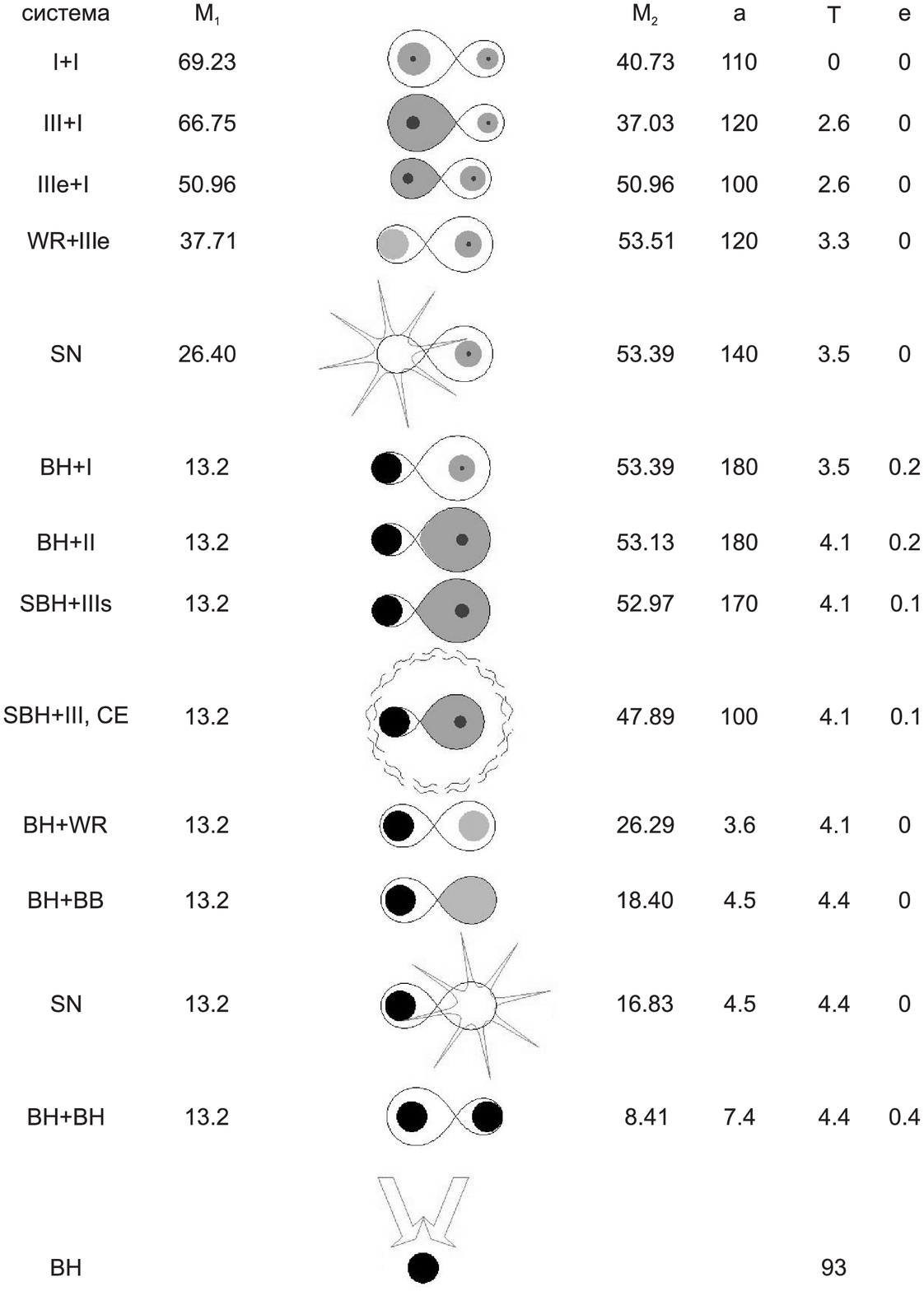}
\vspace{0cm}\caption{Evolutionary scenario for the formation of a Cyg X-3-type system (massive black hole). The notation is described in
Section \ref{results}.}
\label{ris-cyg-x-3-massive}
\end{figure*}

\begin{figure*}[h!]
\hspace{0cm} \epsfxsize=0.95\textwidth\centering
\epsfbox{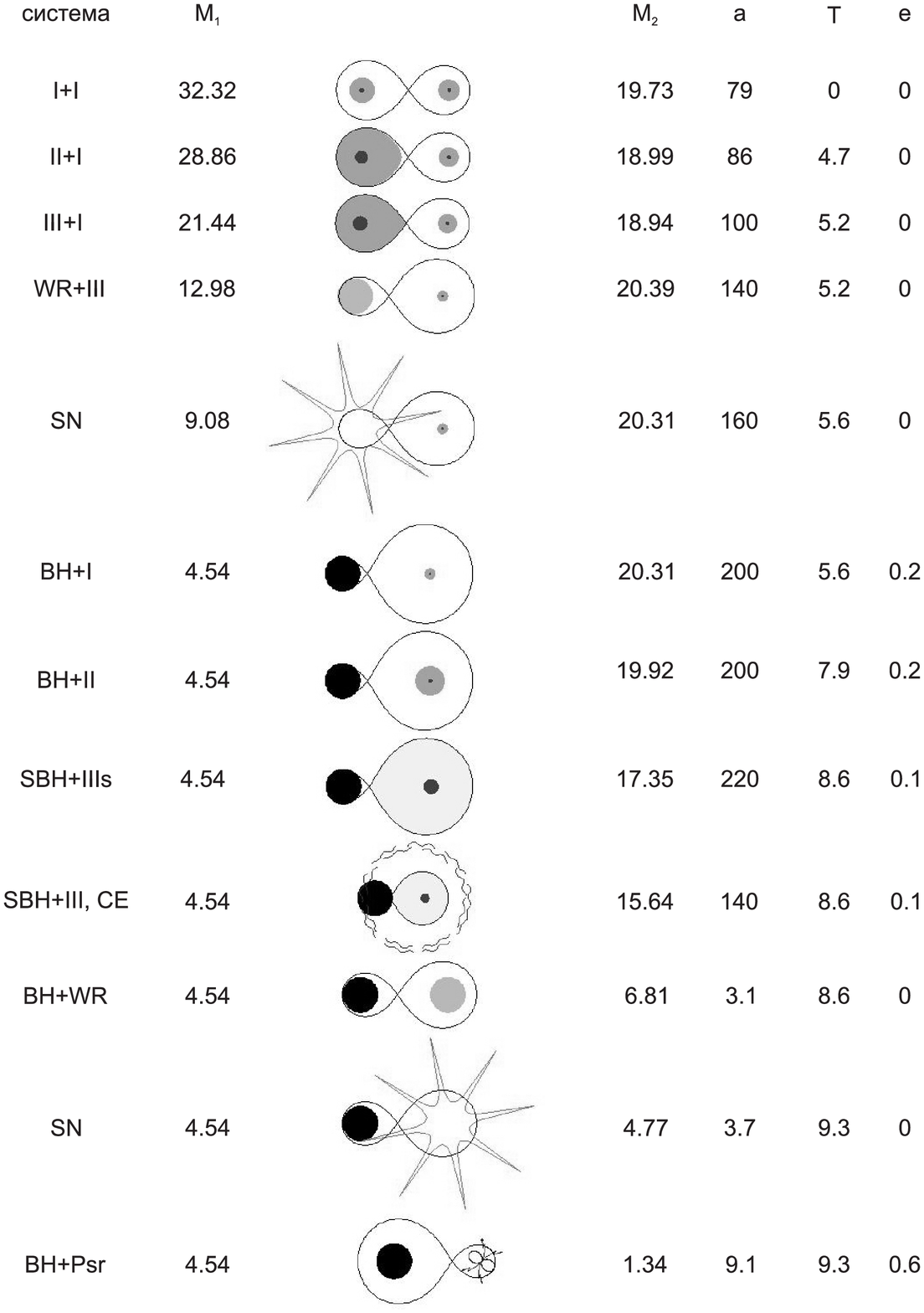} \vspace{0cm}\caption{Evolutionary scenario for the formation of a Cyg X-3-type system (low-mass black hole). The notation is described in
Section \ref{results}.} \label{ris-cyg-x-3-low}
\end{figure*}

\begin{figure*}[h!]
\hspace{0cm} \epsfxsize=0.95\textwidth\centering
\epsfbox{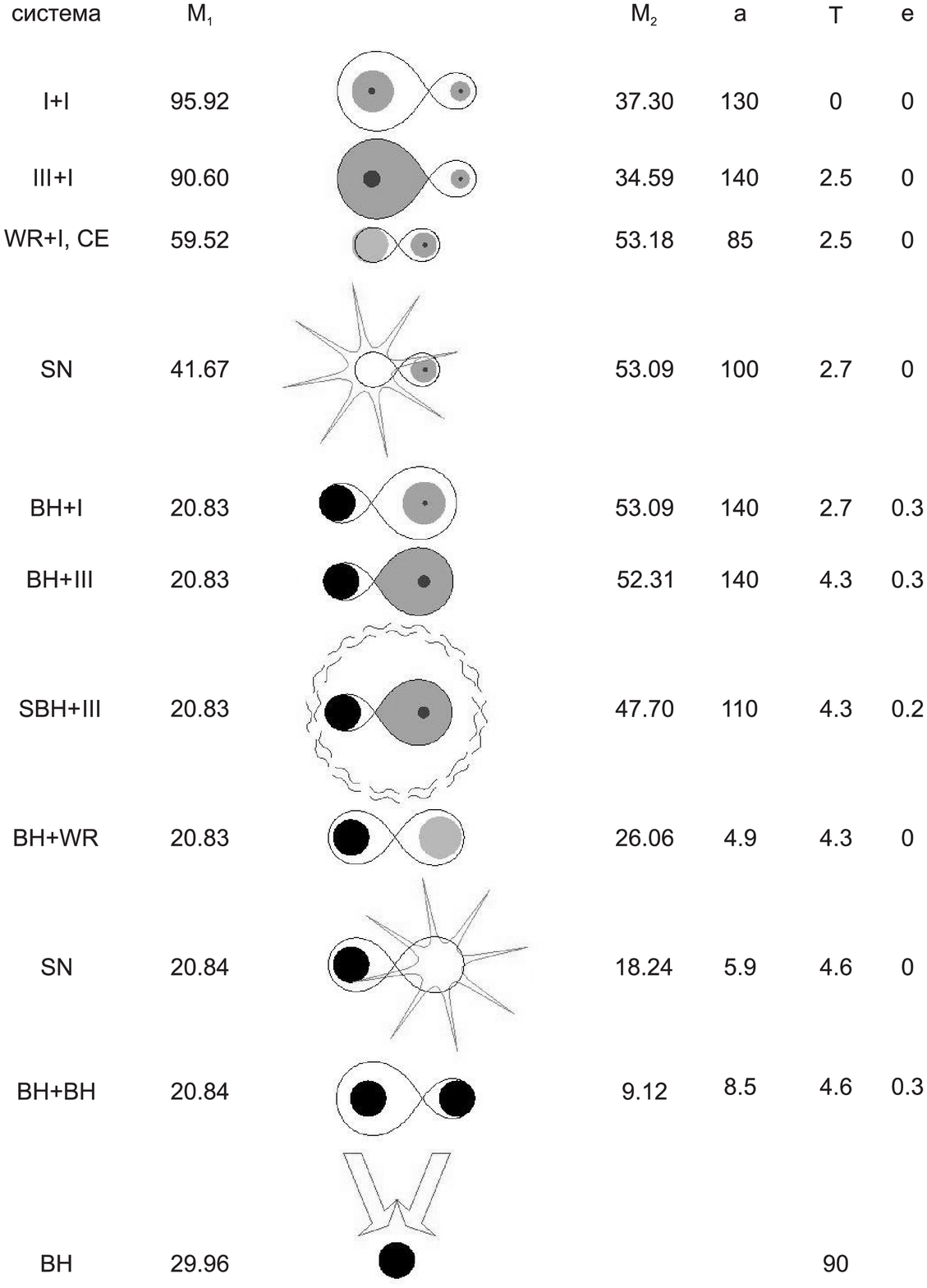} \vspace{0cm}\caption{Evolutionary track for NGC 300 X-1. The notation is described in Section \ref{results}.} \label{ris-ngc-300-x-1}
\end{figure*}

\begin{figure*}[h!]
\hspace{0cm} \epsfxsize=0.95\textwidth\centering
\epsfbox{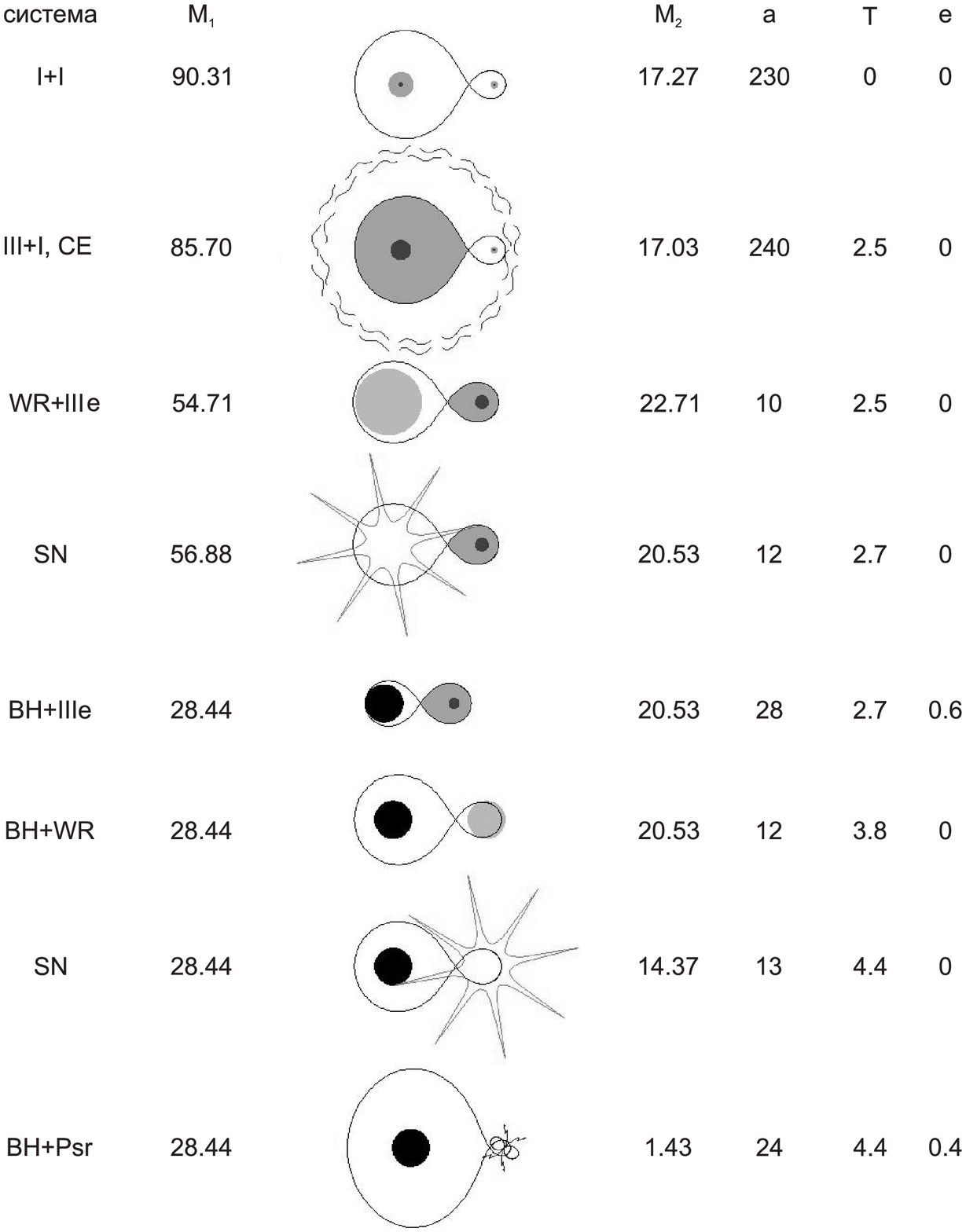} \vspace{0cm}\caption{Evolutionary track for IC 10 X-1. The notation is described in Section \ref{results}.} \label{ris-ic10x1-ns}
\end{figure*}

It follows from Figs. \ref{cygss}-\ref{ris-ic10x1-ns} and Tables
\ref{tr1}-\ref{tr3} that the
origins of Cyg X-3, IC 10 X-1, NGC 300 X-1, and
SS 433 can be described using classical evolutionary
scenarios. The main difference between classical
evolutionary scenarios and the scenarios presented,
for example, in \cite{mink2009}-\cite{bulik2011}\footnote{The evolution of M33 X-7 and IC 10 X-1 was studied in \cite{mink2009} of M33 X-7 in \cite{valsecchi2010} and of IC 10 X-1 and NGC 300 X-1
in \cite{bulik2011}.}, is a much lower rate of
mass loss by massive stars via their stellar winds. It
is shown in \cite{cherepashchuk1984} that the winds of massive stars may
be clumpy. This means that observational estimates
of the wind mass loss rates should be reduced by a
factor of three to five. In addition, the most massive
of the systems considered (M33 X-7, IC 10 X-1, and
NGC 300 X-1) reside in galaxies with lower metallicities
than the Milky Way (see, e.g., \cite{orosz2007} for M33,
\cite{massey2007} for IC 10, and \cite{bresolin2009} for NGC 300).
Decreasing the metallicity also means decreasing the mass loss
rate. Thus, classical evolutionary scenarios with low
mass loss rates, including via stellar winds of massive
stars, have a fairly strong physical basis.

The evolution of Cyg X-3 was studied by Zdziarski
et al. \cite{zdziarski2013} who concluded that the degenerate component
of Cyg X-3 is either a low mass black hole (due
to the accretion-induced collapse of a neutron star
or the collapse of a stellar core during its supernova
explosion) or a neutron star. In the theory underlying
the Scenario Machine, a neutron star cannot
be manifest as an accretor if it is in close pair with
a Wolf-Rayet star \cite{lipunov1982a}.
In addition, as a rule, an
accreting neutron star may accumulate enough mass
to collapse into a black hole only if its companion is
a long lived, low mass star \cite{bogomazov2005}. The precursors of
Wolf-Rayet stars cannot play this role.

We have considered two sets of the main parameters for Cyg X-3-type systems: corresponding to high-mass and low-mass black holes. Observational data supporting the low mass option are provided in \cite{zdziarski2013}. However, the results of calculations carried out in our current study contradict the conclusions of \cite{zdziarski2013}. In particular, the low-mass option implies that the X-ray luminosity of the system generated by the physical mechanisms assumed in the ``Scenario Machine'' should be a factor of ten lower than the real luminosity of Cyg X-3. At the same time, the number of low-mass Cyg X-3-type systems in the Galaxy should be $\sim 10$. If our code is incorrectly estimating the X-ray luminosity, this would suggest that nine out of ten very bright X-ray sources in the Galaxy had not been discovered, which seems implausible, in view of the data provided by the ``Integral'' space observatory and many other observations. Thus, our calculations predict the existence of up to $\sim 10$ systems consisting of a Wolf-Rayet star and a fairly low mass black hole with X-ray luminosities $\lesssim 10^{37}$ erg/s. Apparently, Cyg X-3 must be fairly massive, suggesting that the
high mass option for the evolutionary path of the system is more realistic than the low mass one.

This study was supported by the Russian Foundation for Basic Research (project 12-02-31301 mol-a).

\newpage

\begin{table}
\caption{Evolutionary track for SS 433 most closely describing
the evolution of the system. For notation, see Section \ref{results}.}\label{tr1}
\newcolumntype{C}{>{\centering\arraybackslash}m}
\begin{tabular}{|C{20mm}|C{15mm}|C{15mm}|C{15mm}|C{15mm}|C{15mm}|}
\hline
Íàçâàíèå & $M_1$ &  $M_2$ & $a$ & $T$ & $e$ \\
\hline
I+I     & 66.71 & 38.94 & 31 & 0   & 0 \\
III+I     & 64.43 & 35.65 & 33 & 2.7 & 0 \\
IIIe+I    & 50.04 & 50.04 & 27 & 2.7 & 0 \\
WR+I    & 35.80 & 60.34 & 33 & 3.3 & 0 \\
WR+III    & 35.66 & 60.34 & 33 & 3.3 & 0 \\
WR+IIIe   & 44.33 & 44.33 & 30 & 3.3 & 0 \\
SN      & 45.03 & 39.70 & 32 & 3.6 & 0 \\
SH+III    & 22.51 & 39.70 & 50 & 3.6 & 0.36 \\
BH+WR   & 22.51 & 31.11 & 33 & 3.6 & 0.05 \\
SN      & 22.51 & 21.78 & 40 & 3.9 & 0.05 \\
BH+BH   & 22.51 & 10.89 & 62 & 3.9 & 0.37 \\
\hline
\end{tabular}
\end{table}

\begin{table}
\caption{Additional evolutionary track for an SS 433 type
system. For notation, see Section \ref{results}.}\label{tr2}
\newcolumntype{C}{>{\centering\arraybackslash}m}
\begin{tabular}{|C{20mm}|C{15mm}|C{15mm}|C{15mm}|C{15mm}|C{15mm}|}
\hline
Íàçâàíèå & $M_1$ &  $M_2$ & $a$ & $T$ & $e$ \\
\hline
I+I     & 38.71 & 27.87 & 89 & 0    & 0 \\
II+I    & 33.63 & 25.86 & 99 & 4.2  & 0 \\
III+I     & 33.50 & 25.86 & 99 & 4.2  & 0 \\
IIIE+I    & 27.06 & 27.06 & 100 & 4.2 & 0 \\
WR+I    & 16.71 & 29.66 & 150 & 4.2 & 0 \\
BB+I    & 11.70 & 29.44 & 170 & 4.6 & 0 \\
SN      & 10.60 & 29.55 & 200 & 4.6 & 0 \\
BH+I    & 5.30  & 29.55 & 230 & 4.6 & 0.15 \\
BH+II   & 5.30  & 29.09 & 240 & 5.4 & 0.15 \\
SH+IIIs   & 5.30  & 18.33 & 340 & 5.9 & 0 \\
CE    & 5.30  & 16.46 & 220 & 5.9 & 0 \\
BH+WR   & 5.30  & 11.51 & 180 & 5.9 & 0 \\
SN      & 5.30  & 8.06  & 220 & 6.4 & 0 \\
BH+BH   & 5.30  & 4.03  & 400 & 6.4 & 0.43 \\
\hline
\end{tabular}
\end{table}

\begin{table}
\caption{Additional evolutionary track for an SS 433 type
system. For notation, see Section \ref{results}.}\label{tr3}
\newcolumntype{C}{>{\centering\arraybackslash}m}
\begin{tabular}{|C{20mm}|C{15mm}|C{15mm}|C{15mm}|C{15mm}|C{15mm}|}
\hline
Íàçâàíèå & $M_1$ &  $M_2$ & $a$ & $T$ & $e$ \\
\hline
I+I     & 60.38 & 42.86 & 280 & 0.0 & 0.00 \\
II+I    & 58.10 & 37.31 & 300 & 3.4 & 0.00 \\
III+I     & 52.72 & 36.94 & 320 & 3.8 & 0.00 \\
WR+I    & 31.13 & 46.77 & 380 & 3.8 & 0.00 \\
WR+II   & 24.90 & 46.40 & 410 & 3.9 & 0.00 \\
SN      & 22.42 & 40.58 & 470 & 4.0 & 0.00 \\
BH+II   & 11.21 & 40.58 & 600 & 4.0 & 0.22 \\
SH+IIIs   & 11.21 & 26.22 & 800 & 4.3 & 0.15 \\
CE    & 11.21 & 24.30 & 700 & 4.3 & 0.14 \\
BH+WR   & 11.21 & 21.78 & 190 & 4.3 & 0.00 \\
SN      & 11.21 & 15.24 & 230 & 4.7 & 0.00 \\
BH+BH   & 11.21 & 7.62  & 390 & 4.7 & 0.41 \\
\hline
\end{tabular}
\end{table}

\newpage

\newpage

\newpage

\newpage

\end{document}